\documentclass[twocolumn,preprintnumbers, amssymb,amsmath,aps,floatfix,prl,nofootinbib,superscriptaddress]{revtex4-1}

\usepackage{hyperref}
\usepackage[all]{hypcap}

\usepackage{graphicx}
\usepackage{amsmath,amsthm,amssymb}

\usepackage{color}

\begin{document}

\title{Probing Transverse Momentum Broadening via Dihadron and Hadron-jet Angular Correlations in Relativistic Heavy-ion Collisions}

\author{Lin Chen}
\author{Guang-You Qin}

\author{Shu-Yi Wei}

\author{Bo-Wen Xiao}

\author{Han-Zhong Zhang}

\affiliation{Key Laboratory of Quark and Lepton Physics (MOE) and Institute
of Particle Physics, Central China Normal University, Wuhan 430079, China}

\begin{abstract}
Dijet, dihadron, hadron-jet angular correlations have been reckoned as important probes of the transverse momentum broadening effects in relativistic nuclear collisions.
When a pair of high-energy jets created in hard collisions traverse the quark-gluon plasma produced in heavy-ion collisions, they become de-correlated due to the vacuum soft gluon radiation associated with the Sudakov logarithms and the medium-induced transverse momentum broadening.
For the first time, we employ the systematical resummation formalism and establish a baseline calculation to describe the dihadron and hadron-jet angular correlation data in $pp$ and peripheral $AA$ collisions where the medium effect is negligible.
We demonstrate that the medium-induced broadening $\langle p_\perp^2\rangle$ and the so-called jet quenching parameter $\hat q$ can be extracted from the angular de-correlations observed in $AA$ collisions.
A global $\chi^2$ analysis of dihadron and hadron-jet angular correlation data renders the best fit $\langle p_\perp^2 \rangle \sim 13~\textrm{GeV}^2$ for a quark jet at RHIC top energy.
Further experimental and theoretical efforts along the direction of this work shall significantly advance the quantitative understanding of transverse momentum broadening and help us acquire unprecedented knowledge of jet quenching parameter in relativistic heavy-ion collisions.
\end{abstract}
\maketitle

\textit{Introduction} --- The strongly-coupled quark-gluon plasma (QGP) was one of the most significant discoveries of the Relativistic Heavy Ion Collider (RHIC) experiments, and later confirmed by heavy-ion experiments at the Large Hadron Collider (LHC) \cite{Gyulassy:2004zy, Muller:2012zq}.
Jet quenching, mainly characterized by parton energy loss and transverse momentum broadening, is one of the most important signatures of QGP \cite{Wang:1991xy, Qin:2015srf}.
When a high-energy parton traverses the hot QGP, it interacts with the medium and radiates multiple gluons which take away a significant amount of the energy from the propagating parton \cite{Gyulassy:1993hr, Baier:1996kr, Baier:1996sk, Baier:1998kq, Zakharov:1996fv, Gyulassy:1999zd, Wiedemann:2000za, Arnold:2002ja, Wang:2001ifa, Armesto:2003jh, Djordjevic:2008iz, Majumder:2009ge, CaronHuot:2010bp}.
As a result, both the energy and transverse momentum of the hard parton get modified.
Studies of the nuclear modifications of high transverse momentum hadrons and full jets have shown large parton energy loss effects at RHIC and the LHC \cite{Adler:2003qi, Bass:2008rv, Aamodt:2010jd, Burke:2013yra, Aad:2010bu, Qin:2010mn, Young:2011qx, Zapp:2012ak, Dai:2012am, Wang:2013cia, Blaizot:2013hx}.
In the Baier-Dokshitzer-Mueller-Peigne-Schiff (BDMPS) approach \cite{Baier:1996kr, Baier:1996sk, Baier:1998kq}, medium-induced energy loss and transverse momentum broadening are related.
Their strengths are controlled by the so-called jet transport coefficient $\hat q = {d\langle p_\perp^2\rangle}/{dL}$, which reflects the density of the QGP and may be interpreted as the transverse momentum squared transferred between jet and medium per unit length.
The first effort to quantitatively extract the value of $\hat q$ has been performed by JET Collaboration via comparing several theoretical calculations with experimental data for single hadron production at RHIC and the LHC \cite{Burke:2013yra}.

Since the start of RHIC \cite{Adler:2002tq}, dijet (dihadron) angular correlation has always been an important experimental observable for jet quenching studies in heavy-ion collisions, because its modification in nucleus-nucleus ($AA$) as compared to proton-proton ($pp$) collisions can help us to access the amount of medium-induced transverse momentum broadening.
However, quantitative understanding of dihadron angular correlations has been lacking due to theoretical difficulties.
Previous studies of dijet, dihadron, photon-jet and photon-hadron correlations in heavy-ion collisions mainly focused on the effect of parton energy loss on the modifications of the transverse momentum imbalance distribution and the away-side yield \cite{Qin:2010mn, Zhang:2007ja, Zhang:2009rn, Qin:2009bk}.
Recently, the Sudakov resummation framework has emerged and provided excellent description of dijet angular correlations in both $pp$ and $AA$ collisions at the LHC \cite{Banfi:2008qs, Mueller:2012uf, Mueller:2013wwa, Sun:2014gfa, Mueller:2016gko}.
This allows us to compute angular correlations in high energy collisions by taking into account interesting medium effects, such as the induced transverse momentum broadening, on top of the essential vacuum gluon showers.
The objective of this work is to generalize such resummation framework to perform a systematic study of dihadron and hadron-jet correlations.
We will show that the angular de-correlations of dihadron and hadron-jet pairs can provide a direct probe of medium-induced transverse momentum broadening in relativistic heavy-ion collisions.
Using a global $\chi^2$ analysis of the dihadron and hadron-jet angular correlation data, we further perform a quantitative extraction of jet transverse momentum broadening $\langle p_\perp^2 \rangle$ in QGP.
Compared to the effort by JET Collaboration \cite{Burke:2013yra}, our approach stands as a new and complimentary method for the quantitative understanding of jet transverse momentum broadening and jet quenching parameter in relativistic heavy-ion collisions.

\textit{Implementation of Sudakov factors and medium-induced transverse momentum broadening} --- Our calculation is based on the recent development of the Sudakov resummation in the presence of large medium \cite{Mueller:2016gko, Mueller:2016xoc}. It was found that Sudakov resummation, which takes into account the vacuum parton shower, can be separated from the medium-induced broadening effect in the auxiliary $b$-space.
The physics behind the separation of these two effects is quite similar to the cold nuclear medium case which has been studied in detail in Refs.~\cite{Mueller:2012uf, Mueller:2013wwa}.
In this work, we generalize and apply this Sudakov resummation framework to dihadron and hadron-jet correlations.
For such purpose, we convolute partonic cross sections with the corresponding collinear fragmentation functions together with proper Sudakov factors associated with final state gluon radiations.

We now provide the formalism for calculating dihadron correlation in central rapidity regime in relativistic nuclear collisions; the expressions for dijet and hadron-jet productions are very similar.
The differential cross-section for dihadron production can be cast into \cite{Mueller:2012uf, Mueller:2013wwa, Sun:2014gfa}:
\begin{eqnarray}
\frac{d\sigma}{d\Delta\phi} &=& \sum_{a,b,c,d} \int p_{T}^{h_1} dp_{T}^{h_1} \int p_{T}^{h_2} dp_{T}^{h_2} \int \frac{dz_c}{z_c^2} \int \frac{dz_d}{z_d^2} \nonumber\\
&&\times \int bdb~{J}_0(q_\perp b) e^{-S(Q,b)} x_a f_a(x_a,\mu_b)x_b f_b(x_b,\mu_b) \nonumber\\
&&\times \frac{1}{\pi} \frac{d\sigma_{ab\to cd}}{d\hat{t}} D_c(z_c,\mu_b) D_d(z_d,\mu_b),
\end{eqnarray}
where $J_0$ is the Bessel function of the first kind, $\frac{d\sigma}{d\hat{t}}$ is the leading order partonic cross-section, $x_{a,b}=\textrm{max}(p_{T}^{c}, p_{T}^{d})(e^{\pm y_c}+e^{\pm y_d})/\sqrt{s_{NN}}$ are the momentum fractions carried by the incoming partons from the parent nucleons, $f_{a,b}(x,\mu_b)$ are the parton distribution functions (taken from CTEQ \cite{Pumplin:2002vw}), $z_{c,d}={p_{T}^{h_1,h_2}}/{p_{T}^{c,d}}$ are the momentum fractions carried by the outgoing hadrons from the parent jets, and $D_{c,d}(z, \mu_b)$ are the fragmentation functions (taken from AKK \cite{Albino:2008fy}).
The physical reason which justifies the factorization formula is that different elements receives different one-loop corrections from separated regions in the phase space of the radiated gluon \cite{Mueller:2016xoc}.
According to the derivations in the Sudakov resummation formalism and adopting the so-called $b_{*}$ prescription \cite{Collins:1984kg}, all the factorization scales in the above expression are set to $\mu_b=c_0/b_{*}$ with $c_0\equiv 2e^{-\gamma_E}$ and $b_{*} \equiv b/\sqrt{1+b^2/b_{\max}^2}$.
Here $\mathbf{q}_\perp \equiv \mathbf{p}_{T}^{c} + \mathbf{p}_{T}^{d}$ indicates the dijet transverse momentum imbalance, which originates from the combination of initial and final state vacuum radiations as well as medium-induced transverse momentum broadening. The Sudakov factor $S(Q, b)$ are functions of the hard scale $Q = \sqrt{x_a x_b s_{NN}}$ and the Fourier transform of $q_\perp$ in the auxiliary $b$-space.

Using $b_{*}$ prescription, the contributions from vacuum radiations to the Sudakov factor can be separated into perturbative and non-perturbative parts \cite{Mueller:2012uf, Mueller:2013wwa, Sun:2014gfa}:
\begin{equation}
S_{\textrm{vac}}(Q,b)=S^i_{\textrm{p}}(Q,b)+S^f_{\textrm{p}}(Q,b)+S_{\textrm{np}}(Q,b),
\end{equation}
where $S_{\textrm{p}}^{i}$ ($S_{\textrm{p}}^{f}$) denotes the initial (final) state contribution to the Sudakov factor from the incoming (outgoing) partons, and $S_{\textrm{np}}$ represents the total non-perturbative contributions from both initial and final states.

At one-loop order, the perturbative Sudakov factor $S^i_{\textrm{p}}(Q,b)$ for initial state radiation can be written as:
\begin{equation}
S^i_{\textrm{p}}(Q,b)= \sum_{i=a, b} \int_{\mu_b^2}^{Q^2}\frac{d\mu^2}{\mu^2}\left[A_i\ln\left(\frac{Q^2}{\mu^2}\right)+B_i\right],
\end{equation}
where the coefficients $A_g=C_A\frac{\alpha_s}{2\pi}$, $B_g=-C_A\beta_0\frac{\alpha_s}{\pi}$ for incoming gluons, and $A_q=C_F\frac{\alpha_s}{2\pi}$, $B_q=-\frac{3}{4}C_F\frac{\alpha_s}{\pi}$ for incoming quarks.
The final state perturbative Sudakov factor $S^f_{\textrm{p}}(Q,b)$ depends on jet or hadron production \cite{Mueller:2015ael, Xie:2013cba}.
For jet final state, an additional factor $D \ln\frac{1}{R^2}$ is added along with the $A$, $B$ terms to account for the contribution from the gluon radiation outside the jet cone with size $R$, where $D_g=C_A\frac{\alpha_s}{2\pi}$ for gluon jets and $D_q=C_F\frac{\alpha_s}{2\pi}$ for quark jets.
For hadron production, the final state radiation effects are described by the same $A$, $B$ coefficients as the initial Sudakov factors.
Note that for hadron production at mid-rapidity, only half of the final state Sudakov factors contribute to the azimuthal de-correlation.

The non-perturbative Sudakov factor $S_{\textrm{np}}(Q,b)$ may be contributed from initial and final states. For incoming quark, the non-perturbative factor is given as \cite{Su:2014wpa, Prokudin:2015ysa}:
\begin{equation}
S^q_{\textrm{np}}(Q,b)=\frac{g_1}{2} b^2 +\frac{g_2}{2}\ln\frac{Q}{Q_0}\ln \frac{b}{b_{*}},
\end{equation}
where $g_1 = 0.212$, $g_2 = 0.84$, $Q_0^2=2.4~\textrm{GeV}^2$, and $b_\textrm{max}=1.5~\textrm{GeV}^{-1}$. For incoming gluon, $S^g_{\textrm{np}}(Q,b)= \frac{C_A}{C_F} S^q_{\textrm{np}}(Q,b)$.
Other choices of the non-perturbative part yields similar angular correlation numerically \cite{Landry:2002ix, Sun:2012vc}.
We note that due to the well-known issue of non-universality \cite{Collins:2007nk, Rogers:2010dm} in dijet processes in hadron collisions, the non-perturbative Sudakov factors in dihadron and hadron-jet productions are expected to be different from those used in deep inelastic scattering (DIS) or Drell-Yan (DY) processes.
Being lack of quantitative understanding of the non-universality, we rely on the numerical fit to $pp$ data; our best fit shows that we need to multiply $S_{\textrm{np}}(Q,b)$ as obtained from DY and DIS by a factor of $4.8$ for dihadron and $1.3$ for hadron-jet productions.
Although the non-perturbative Sudakov factor is process-dependent, it could be universal for different momentum cuts and collision energies for a given process.

To include the contributions from the medium-induced transverse momentum broadening to the angular de-correlations, a single resummed formula for the Sudakov factor can be derived \cite{Mueller:2016gko, Mueller:2016xoc}:
\begin{eqnarray}
S_{\textrm{med}}(Q,b) = S_{\textrm{vac}}(Q,b)
+\frac{b^2}{4} \frac{1}{2} \left(\langle p_\perp^2 \rangle_c +\langle p_\perp^2 \rangle_d \right),
\end{eqnarray}
where $\langle p_\perp^2 \rangle_{c,d}$ indicates the amount of the medium-induced broadening experienced by each outgoing jet, after averaging over different paths.
The medium-induced broadening for quark and gluon jets are related as $\langle p_\perp^2\rangle_g = \frac{C_A}{C_F} \langle p_\perp^2\rangle_q$.
The factor ${1}/{2}$ is to account for the fact that at mid-rapidity, only half of the medium broadening effect contributes to the azimuthal angle de-correlation (the other half goes into the beam direction).

\begin{widetext}

\begin{figure}[tbp]
\begin{center}
\includegraphics[width=5.55cm]{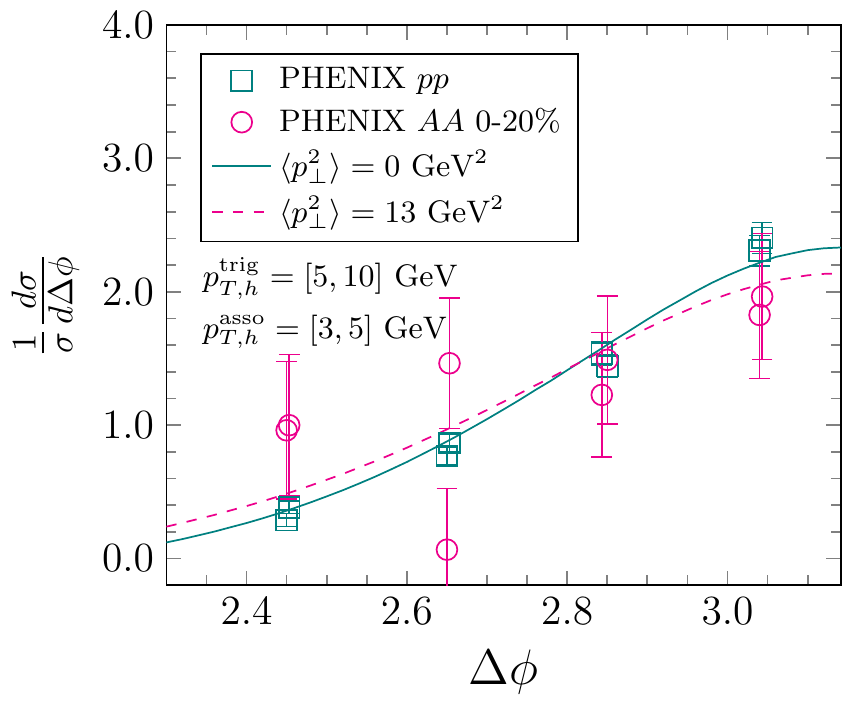}
\includegraphics[width=5.55cm]{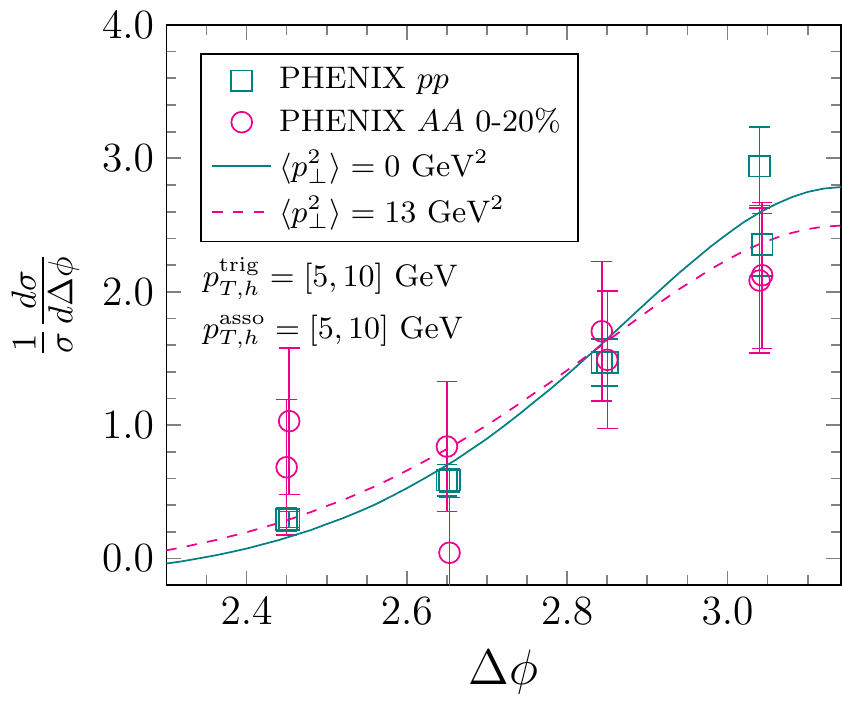}
\includegraphics[width=5.55cm]{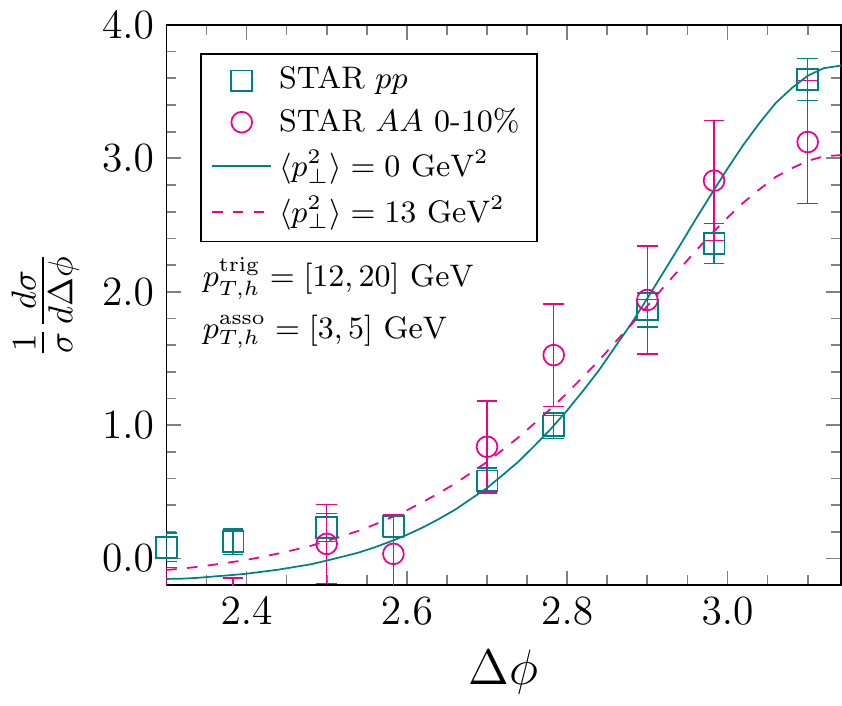}
\end{center}
\vspace{-20pt}
\caption[*]{Normalized dihadron angular correlation compared with PHENIX \cite{Adare:2007vu} and STAR \cite{STAR:2016jdz} data.}
\label{dihadron}
\end{figure}

\begin{figure}[tbp]
\begin{center}
\includegraphics[width=5.55cm]{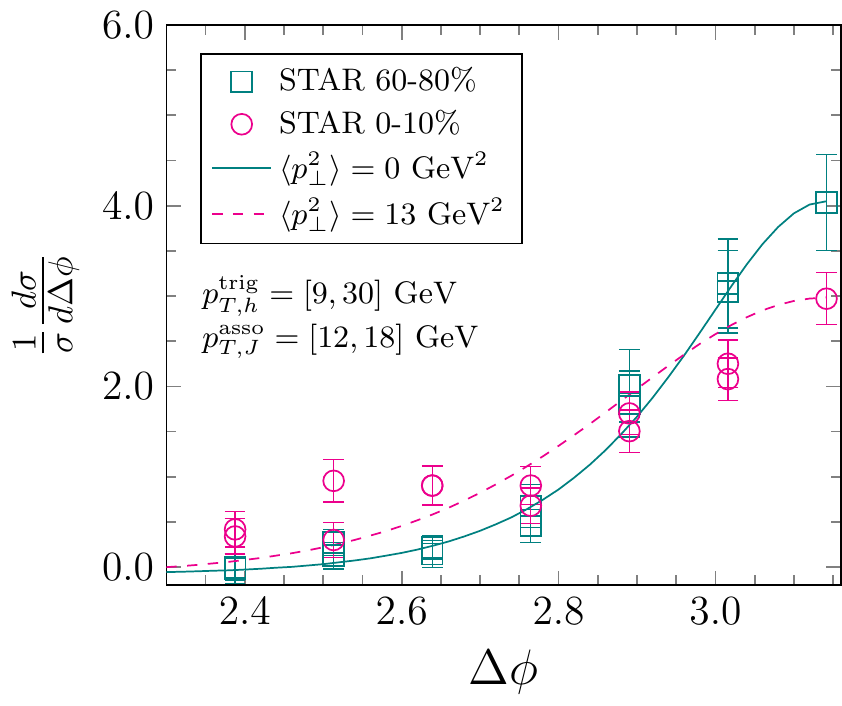}
\includegraphics[width=5.55cm]{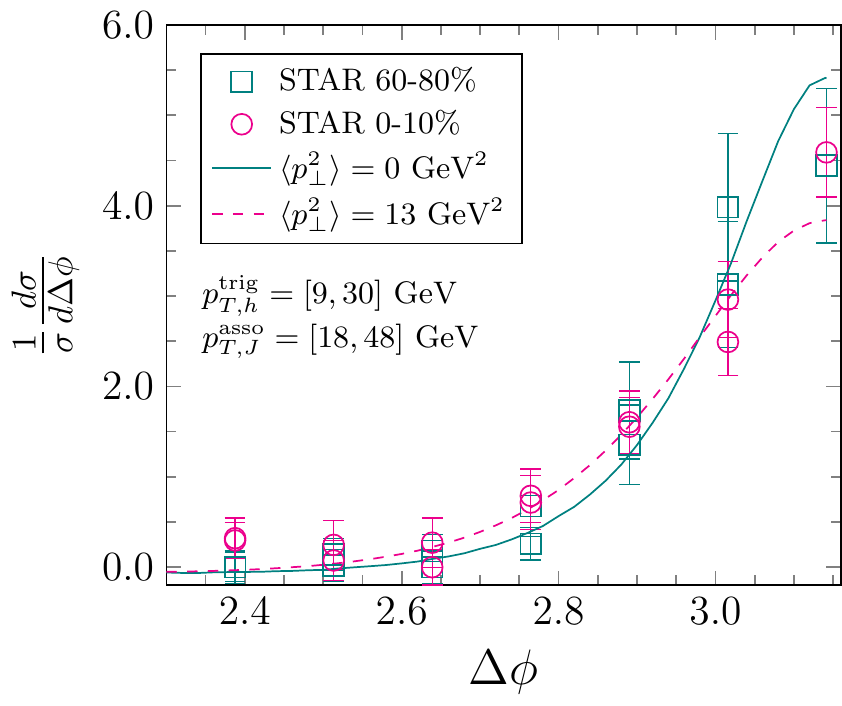}
\includegraphics[width=5.55cm]{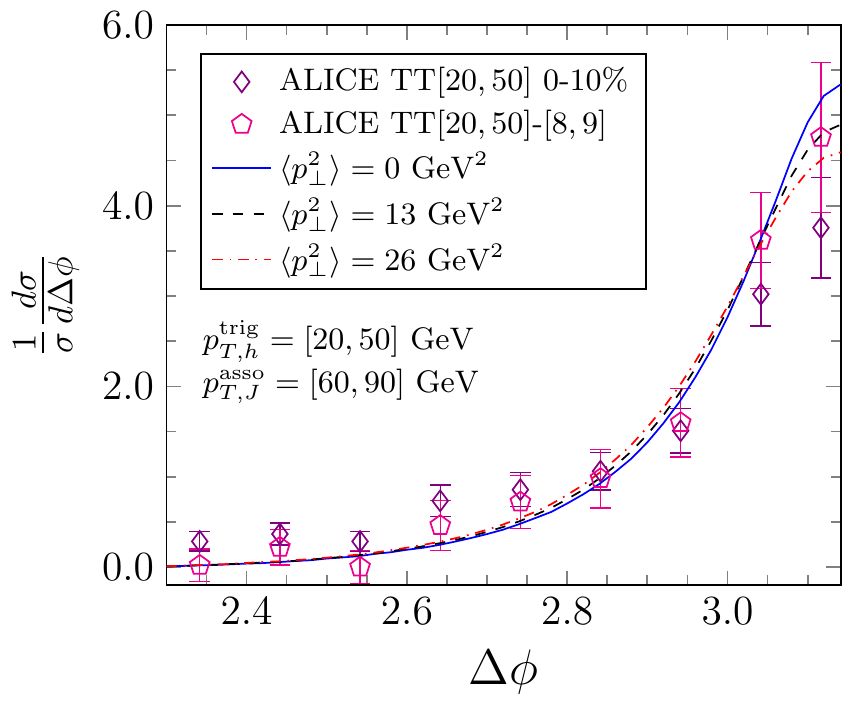}
\end{center}
\caption[*]{Normalized hadron-jet angular correlation compared with STAR \cite{Jacobs:2015srw} and the ALICE \cite{Adam:2015doa} data.
A factor of $3/2$ is multiplied to the charged jet energy for our calculation to account for the energy carried by neutral particles.}
\label{hadron-jet}
\end{figure}

\end{widetext}

\textit{Numerical results} --- Based on the above Sudakov resummation formalism which includes both vacuum and medium-induced transverse broadening effects, we have performed a comprehensive study of the angular correlations of dijets, dihadrons and hadron-jets in both $pp$ and $AA$ collisions at RHIC and the LHC energies with various transverse momentum cuts.

First, we check dijet angular correlation at the LHC for $p_{T}^{\textrm{trig}} > 120~\textrm{GeV}$ following Ref. \cite{Mueller:2016gko}, and find perfect agreement with the CMS data \cite{Chatrchyan:2011sx} (not shown). We also observe that the angular de-correlation is completely insensitive to the medium-induced broadening effects and the non-perturbative Sudakov contributions due to overwhelming perturbative Sudakov contribution for high energy jet production at the LHC. This implies that one needs to look for events with smaller jet energies or in lower energy collisions in order to probe the medium-induced transverse momentum broadening.

We then compute dihadron angular correlation in both $pp$ and $AA$ collisions by convoluting final state partons with the corresponding fragmentation functions together with proper final state Sudakov factors.
Here we focus on dihadrons with the transverse momenta larger than 3-5~GeV to avoid large non-perturbative effects and flow contamination.
Also to minimize the contribution from the intrinsic broadening of fragmentation functions, larger transverse momentum hadrons are preferred.
In Fig. \ref{dihadron}, we show our calculations of dihadron angular correlations compared with PHENIX (charge dihadrons) \cite{Adare:2007vu} and STAR ($\pi^0$ + charge hadron) \cite{STAR:2016jdz} data.
As we are interested in the shape of the angular correlations, all data and theoretical curves are normalized to unity.
One can see, with the perturbative Sudakov resummation and the same setup for non-perturbative Sudakov factor, reasonable descriptions of $pp$ data are obtained for various transverse momentum cuts.
Both data and theoretical calculations show that the transverse momentum broadening contributed from the vacuum radiation tends to flatten the shape of angular correlation.
Using $pp$ calculations as the baseline, we calculate dihadron angular correlation in $AA$ collisions by incorporating the additional broadening effect due to jet-medium interaction.
We find that the amount of medium-induced transverse broadening $\langle p_\perp^2 \rangle \sim 13~\textrm{GeV}^2$ can describe the $AA$ data from PHENIX and STAR.

Very recently, hadron-jet correlations in $AA$ collisions have been measured by both ALICE and STAR Collaborations \cite{Adam:2015doa, Jacobs:2015srw}.
We compute hadron-jet angular correlations by convoluting one final state jet with the corresponding fragmentation function, combined with proper final state Sudakov factors as well as jet cone contribution for the other jet.
Here we use the peripheral $AA$ data as the baseline assuming that the medium effect is negligible there.
As shown in Fig.~\ref{hadron-jet}, hadron-jets measured by the ALICE are not sensitive to medium-induced broadening effect due to tremendous amount of vacuum Sudakov effects.
However, hadron-jets measured by STAR are quite sensitive to the medium effect and can help us to probe the transverse momentum broadening experienced by the hard jets traversing the QGP medium.
The comparison with STAR hadron-jet data gives $\langle p_\perp^2 \rangle \sim 13~\textrm{GeV}^2$, similar to dihadron result.
We mention that STAR and the ALICE data have not yet been corrected for azimuthal smearing due to uncorrelated background.
We also note that for hadron-jets (and dijets), only the radiation outside jet cone contributes to the angular de-correlation. Therefore, by varying the jet size, the medium effects on jet internal structure may be studied as well.

\begin{figure}[tbp]
\begin{center}
\includegraphics[width=7.77cm]{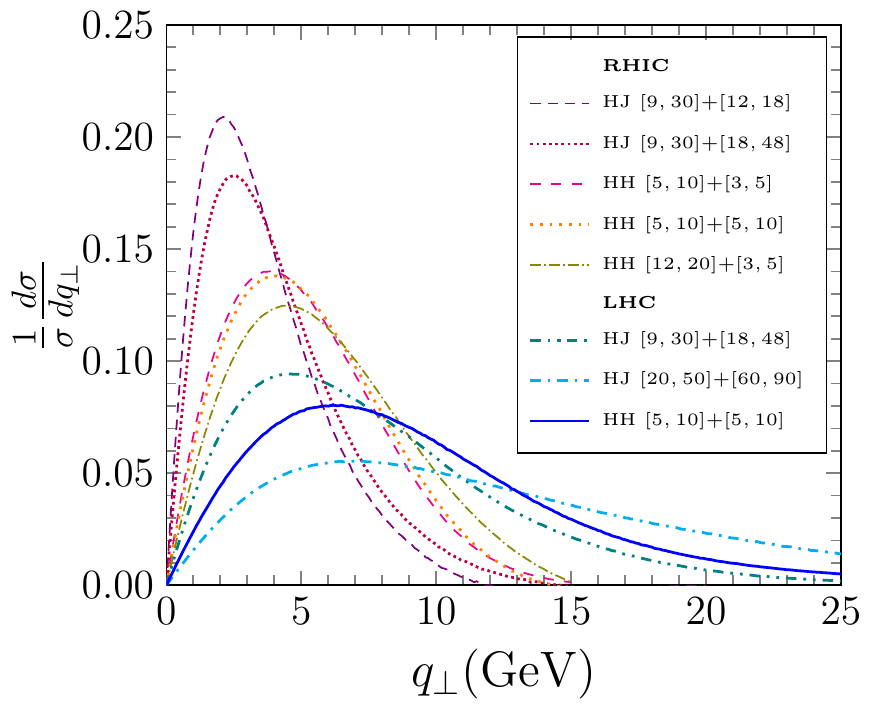}
\end{center}
\caption[*]{Normalized dijet $q_\perp$ distributions for dihadron (HH) and hadron-jet (HJ) correlations in $pp$ collisions for various transverse momentum cuts at RHIC and the LHC.} 
\label{qtdist}
\end{figure}

For a better understanding of the sensitivity of the angular correlations to medium-induced broadening effects, we show in Fig.~\ref{qtdist} the normalized dijet $q_\perp$ distributions for dihadrons and hadron-jets in $pp$ collisions for various transverse momentum cuts at RHIC and the LHC.
The peak value (denoted as $q_\perp^{\ast}$) of the distribution indicates the typical value of dijet transverse momentum imbalance due to the vacuum Sudakov effect.
Using the pocket formula $q_{\perp \textrm{AA}}^{\ast 2}\simeq q_{\perp \textrm{pp}}^{\ast 2} + \langle p_\perp^2 \rangle$, we can see that smaller transverse momentum jets are in general more effective in probing the medium-induced broadening effect via dihadron and hadron-jet angular correlations due to smaller vacuum Sudakov effect.

Finally, we perform a global $\chi^2$ analysis by combining dihadron and hadron-jet angular correlation data from STAR (PHENIX dihadron data are not included in the analysis due to large uncertainties).
Here we focus on the region $|\pi-\Delta \phi| < \pi/4$ to suppress the contribution from the rare hard processes.
Using CERN MINUIT package \cite{James:1975dr}, our $\chi^2$ analysis yields the transverse momentum broadening $\langle p_\perp^2 \rangle = 13^{+5}_{-4}~\textrm{GeV}^2$ for a quark jet in central Au-Au collisions at 200A~GeV at RHIC.
Note that our resummation formalism includes the contributions up to next-to-leading logrithmic (NLL) accuracy; the quoted ($1\sigma$) errors do not include the uncertainties from the high-order corrections beyond NLL. 

To directly compare to the JET result \cite{Burke:2013yra}, we further extract the jet quenching parameter $\hat{q}$ by convoluting the angular distribution 
with a realistic modeling of the collision geometry and the initial jet production locations from the Glauber model, 
and the space-time evolution of the medium simulated via the OSU (2+1)-dimensional viscous hydrodynamics code \cite{Song:2007ux, Qiu:2011hf}. 
Considering the radiative correction to the medium-induced broadening $\langle p_\perp^2 \rangle$ and $\hat{q}$ \cite{Wu:2011kc,Liou:2013qya,Kang:2013raa,Iancu:2014kga, Blaizot:2014bha}, we take the leading double logarithmic resummed expression from Ref. \cite{Liou:2013qya}:
$\langle p_\perp^2 \rangle = \hat{q} L {I_1\left[2\sqrt{\bar{\alpha}_s}\ln\left({L^2}/{l_0^2}\right)\right]}/\left[{\sqrt{\bar{\alpha}_s} \ln\left({L^2}/{l_0^2}\right)}\right]$,
with $\bar{\alpha}_s = {\alpha_s N_c}/{4\pi}$, $I_1$ the Bessel function of the second kind, and $\l_0$ the size of the medium constituents (we use $\alpha_s = 1/3$ and $l_0=0.3~\textrm{fm}$ in the calculation).
As a common practice, we relate the leading-order $\hat{q}$ to the medium temperature as $\hat{q} \propto T^3$. 
The $\chi^2$ analysis with MINUIT package renders $\hat{q}_0 = 3.9^{+1.5}_{-1.2}~\textrm{GeV}^2/\textrm{fm}$ for a quark jet at $\tau_0 = 0.6~\textrm{fm}/\textrm{c}$ (the starting time of hydrodynamics) at the center of the medium produced in central Au-Au collisions at 200A~GeV at RHIC. 
Our central value is a few times larger than (but not inconsistent with) the value $\hat{q}_0 =1.2^{+0.3}_{-0.3}~\textrm{GeV}^2/\textrm{fm}$ for a $10$~GeV quark jet extracted by JET Collaboration by comparing with the nuclear modification of single hadron production \cite{Burke:2013yra}.
We also estimate the contribution from jet energy loss and find that its influence on the angular distribution is weak. 
However, due to the flatness around the minimum $\chi^2$ value mainly originating from the large error bars in the data, a 5-10\% fractional energy loss on the associated jets leads to the increase of the central $\hat{q}$ value by 15-35\%. 
A detailed analysis of jet energy loss and its effect on the nuclear modifications of dihadron and hadron-jet angular correlations as well as yields is left to a future study.

\textit{Discussion and conclusion} --- 
To conclude, we have applied the Sudakov resummation formalism, including the contribution from medium-induced transverse momentum broadening effect, to compute dijet, dihadron and hadron-jet angular correlations in $pp$ and $AA$ collisions at RHIC and the LHC.
We have demonstrated that the shape of the angular correlations can provide a new gateway to quantify the medium-induced broadening experienced by the hard jets.
The comparison with RHIC dihadron and hadron-jet angular de-correlation data allows us to extract the medium-induced broadening $\langle p_\perp^2\rangle \sim 13~\textrm{GeV}^2$ for a quark jet at RHIC top energy.
We believe that with more precise experimental data on dijet, dihadron and hadron-jet angular correlations, one is able to acquire more precise understanding of jet quenching parameter and the properties of hot QCD medium created in relativistic heavy-ion collisions.

We note that our framework can be applied to study the nuclear modifications of dijet (dihadron) momentum imbalance distributions and the triggered away-side yield.
It can also be generalized to calculate dihadron productions in forward rapidity region and study the gluon saturation effect.
As a final remark, dihadron and hadron-jet correlations for RHIC beam energy scan may serve as an additional indicator for the formation of QGP medium due to their sensitivity to the medium broadening effect while $\hat q$ of cold nuclear matter \cite{Baier:1996sk, Wang:2009qb} is at least an order of magnitude smaller than that of the hot nuclear medium.

\textit{Acknowledgment} --- We thank P. Jacobs, Y.-X. Mao, A. H. Mueller, J.-W. Qiu, A. Schmah, X.-N. Wang and F. Yuan for interesting discussions and comments.
This work is supported by the NSFC under Grant Nos. 11375072, 11435004 and 11575070, and the Major State Basic Research Development Program in China under Contract No. 2014CB845400.

\bibliographystyle{h-physrev5}
\bibliography{refs_qhat}

\end{document}